\documentclass[a4paper,11pt]{article}

\usepackage{pos_nectarcam}
\usepackage{lipsum} 

\title{Performances of an upgraded front-end-board for the NectarCAM camera}

\ShortTitle{Performances of an upgraded front-end-board for the NectarCAM camera}

\author*[1]{Federica Bradascio}
\author[1]{F.~Brun}
\author[2]{F.~Cangemi}
\author[3]{S.~Caroff}
\author[1]{E.~Delagnes}
\author[4]{D.~Gascon}
\author[1]{J.-F.~Glicenstein}
\author[5]{C.~Juramy-Gilles}
\author[5]{J.-P.~Lenain}
\author[5]{J.-L.~Meunier}
\author[4]{A.~Sanuy} 
\author[1]{P.~Sizun}
\author[5]{F.~Toussenel}
\author[1]{B.~Vallage}
\author[5]{V.~Voisin}

\affiliation[1]{IRFU, CEA, Universit\'e Paris-Saclay, F-91191 Gif-sur-Yvette, France}

\affiliation[2]{Universit\'e de Paris, CNRS, Astroparticule et Cosmologie, F-75013 Paris, France}
\affiliation[3]{Univ. Savoie Mont Blanc, CNRS, Laboratoire d'Annecy de Physique des Particules - IN2P3, 74000 Annecy, France}
\affiliation[4]{Departament de F\'isica Qu\`antica i Astrof\'isica, Institut de Ci\`encies del Cosmos, Universitat de Barcelona, IEEC-UB, Mart\'i i Franqu\`es, 1, 08028, Barcelona, Spain}
\affiliation[5]{Sorbonne Universit\'e, CNRS/IN2P3, Laboratoire de Physique Nucl\'eaire et de Hautes Energies, LPNHE, 4 place Jussieu, 75005 Paris, France}

\forColl{CTA NectarCAM} 
\emailAdd{federica.bradascio@cea.fr}

\abstract{

The Front-End Board (FEB) is a key component of the NectarCAM camera, which has been developed for the Medium-Sized-Telescopes (MST) of the Cherenkov Telescope Array Observatory (CTAO). The FEB is responsible for reading and converting the signals from the camera's photo-multiplier tubes (PMTs) into digital data, as well as generating module level trigger signals. This contribution provides an overview of the design and performances of a new version of the FEB that utilizes an improved version of the NECTAr chip. The NECTAr chip includes a switched capacitor array for sampling signals at 1 GHz, and a 12-bit analog-to-digital converter (ADC) for digitizing each sample when the trigger signal is received. The integration of this advanced NECTAr chip significantly reduces the deadtime of NectarCAM by an order of magnitude as compared to the previous version. This contribution also presents the results of laboratory testing of the new FEB, including measurements of timing performance, linearity, dynamic range, and deadtime.

\ConferenceLogo{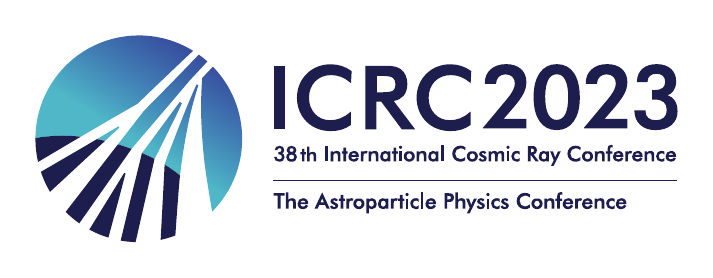}

\FullConference{The 38th International Cosmic Ray Conference (ICRC2023)\\ 26 July -- 3 August, 2023\\ Nagoya, Japan}
}

\begin{document}

\maketitle

\section{Introduction}\label{sec1}
The Cherenkov Telescope Array (CTA) is a planned Imaging Atmospheric Cherenkov Telescope (IACT) designed to detect very high-energy (VHE) gamma-ray photons ranging from a few tens of GeV to above 100 TeV~\cite{CTAconcept}. VHE gamma rays are indirectly observed by detecting the Cherenkov light produced when these photons interact in the Earth's atmosphere, creating particle cascades. The CTA observatory (CTAO), with its tenfold increase in sensitivity, will outperform current generation IACT arrays significantly. It will consist of over 60 telescopes distributed across two sites: one in La Palma, Spain (Northern Hemisphere), and the other in Chile (Southern Hemisphere), providing access to nearly the entire night sky. To cover the wide energy range, the CTA employs three types of Cherenkov telescopes: Small-Sized Telescopes (SSTs), Medium-Sized Telescopes (MSTs), and Large-Sized Telescopes (LSTs). The MST telescopes at the CTA-North site will be equipped with the NectarCAM camera~\cite{nectarcam}.

The NectarCAM camera has a modular design, with a basic element (``module") consisting of a focal plane module (FPM) and a front-end board (FEB). The FPM  \cite{2021NIMPA100765413T} is composed of seven 1.5" R12992-100-05 Hamamatsu photomultiplier tubes (PMTs)  associated with high voltage and pre-amplification boards (HVPA), and equipped with Winston cone light concentrators \cite{lightconcentrators}. The Cherenkov light impinging on the camera is first detected in the focal plane, converted into an electric signal by the PMTs and preamplified towards two gain channels. 
The signal is then amplified a second time in the FEB by a custom amplifier called ACTA, where it is split into three channels: low and high gain channels, and trigger channel. 

At the core of the sampling and digitization process of NectarCAM is the NECTAr chip~\cite{Nectar2012, nectar0}. This chip is a switch capacitor array that can sample signals at 1 GHz and is coupled with a 12-bit analog-to-digital converter (ADC). Acting as a circular buffer, the NECTAr chip stores data until a camera trigger is activated.
Currently, the readout of the NECTAr chip significantly contributes to the deadtime of NectarCAM. To address this issue, a new version of the NECTAr chip and the front-end board (FEB version 6, hereafter FEBv6) has been developed for the first on-site deployment of NectarCAM.
The key improvement is the implementation of a "ping-pong" operation, where the analogue memory is divided into two sub-arrays of 512 cells each. This division allows for simultaneous read and write access, significantly reducing the deadtime associated with data acquisition.
In the new NECTAr chip, the signal is continuously written into one sub-array while the other sub-array is frozen and digitized. When a trigger occurs, the frozen sub-array is read while the write operation switches to the other sub-array. This alternating operation ensures a continuous flow of data without the need to wait for the readout process to complete before writing new data.
This approach reduces the camera's minimum deadtime by an order of magnitude, decreasing it from $7~\mu$s to $0.7~\mu$s. Consequently, the deadtime at a trigger rate of 7~kHz is reduced from 5.2\% to 0.5\%.
By achieving an order of magnitude reduction in deadtime, the NectarCAM camera will be capable of operating at a higher trigger rate and lower the energy threshold to around 50 GeV when operated at a trigger rate of 7 kHz~\cite{trigger}. 

In this contribution, we present the results of three verification tests conducted at the integration facility in IRFU, CEA Paris-Saclay (France), performed in a darkroom using 10 new FEBv6 modules equipped with the upgraded NECTAr chips.

\section{Pixel timing precision}
We conducted measurements to determine the timing precision of the pixels in the 10 FEBv6, following the procedure outlined in \cite{timing_paper}. By employing a laser source with a frequency of 1 kHz and varying pulsed energies ranging from 8 pJ to 20 pJ, we measured the position of the maximum sample within the 60 ns readout window (referred to as Time of Maximum or TOM) for each photon pulse in every pixel.

\begin{figure}[t!]
    \centering
    \includegraphics[width=0.9\columnwidth]{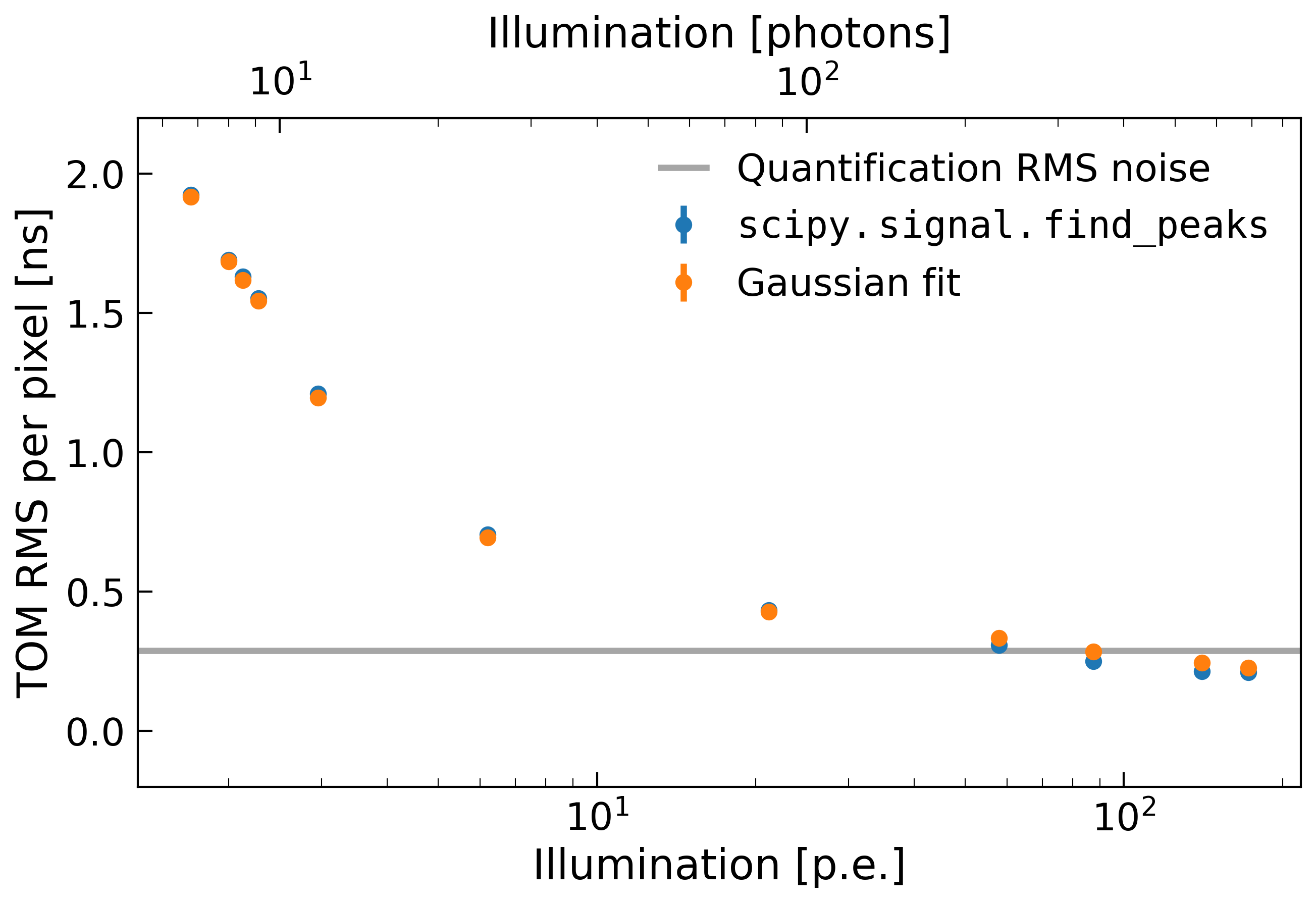}
    \caption{Timing precision per pixel (in ns) as a function of the charge of the illumination signal (in photons and photoelectrons on the bottom and top of the horizontal axis, respectively). The timing resolution is given by the mean of the RMS distribution over all the 70 pixels. The 2 methods described in \cite{timing_paper} are shown (in blue and orange). The gray solid line shows the quantization (RMS) noise given by $\frac{1}{\sqrt{12}}$~ns. }
    \label{fig:pixel_res}
\end{figure}

The TOM values were obtained using the two methods described in \cite{timing_paper}. To estimate the systematic timing uncertainty for each pixel, we calculated the root mean square (RMS) of the TOM distributions, taking into account the contribution of the random starting time of the NECTAr chip readout. \autoref{fig:pixel_res} illustrates the weighted mean of the RMS values for all pixels, considering the illumination charge, using both methods. The weights were determined based on the inverse square of the standard deviation of the TOM distribution for each pixel. For incident light intensities above approximately 10 photons, both methods indicate a pixel precision below 1 ns. However, at an illumination charge of 900 p.e., as indicated by the solid gray line, the precision becomes limited by the quantization RMS noise of the chip responsible for signal readout. This limitation arises from the quantization of the trigger acceptance signal caused by the asynchronous arrival of the trigger in relation to the NECTAr chip clock. Given that the NECTAr chip samples at 1 GHz, this corresponds to an RMS value of $1/\sqrt{12}$ ns, which is approximately 290 ps.
    
\section{Deadtime}

The reduction of deadtime was evaluated by measuring the shortest time interval, denoted as $\Delta t$, between two consecutive events while illuminating the camera with various random sources. Three random sources were utilized: a night-sky background source, a flat-field calibration source (FFCLS) and a laser both triggered by a random generator\footnote{The random generator consists of a photon source triggering a photomultiplier \cite{timing_paper}.}.

\begin{figure}[t]
    \centering
    \includegraphics[width=0.9\columnwidth]{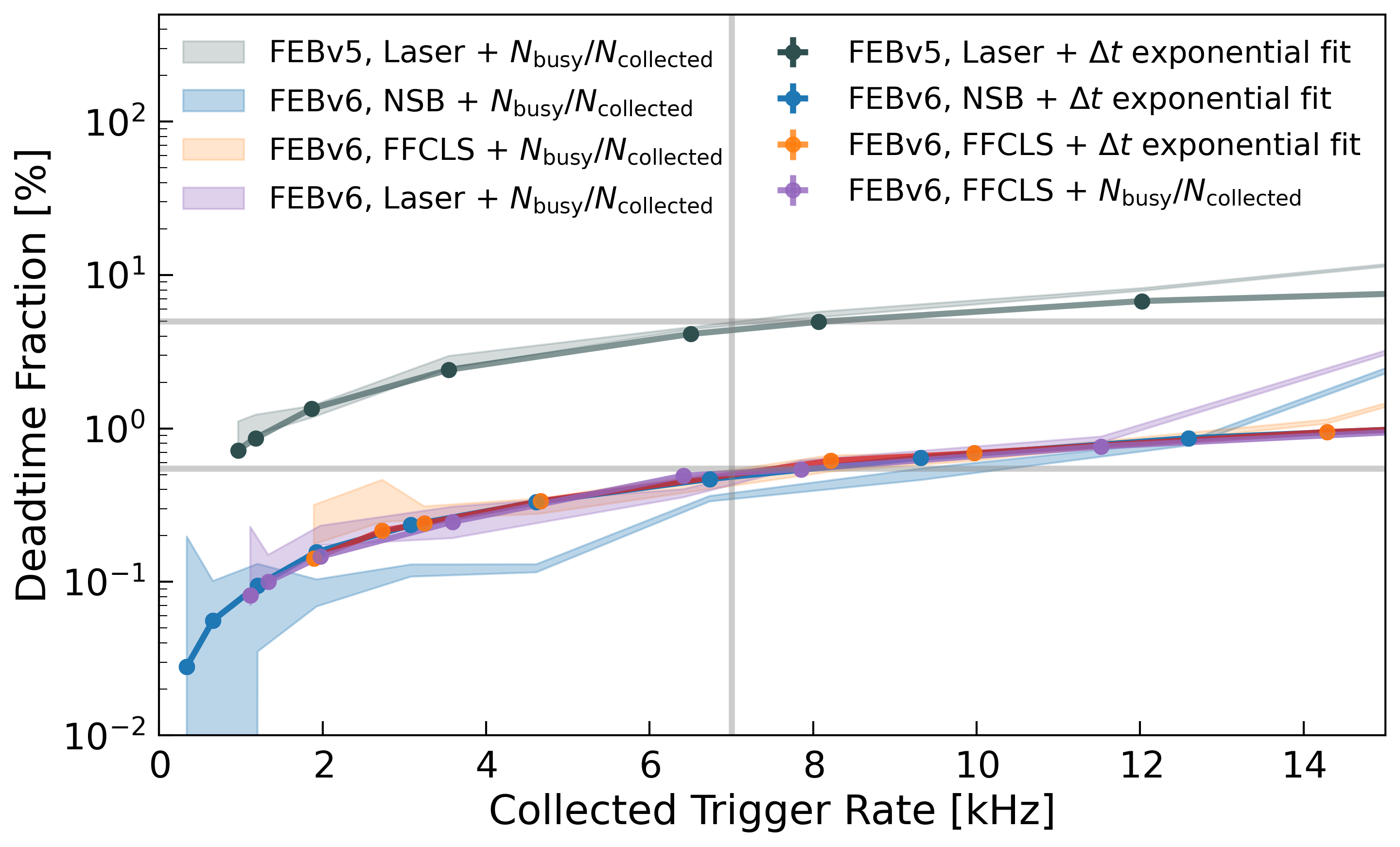}
    \caption{Deadtime fraction for the 10 FEBv6 (in color) and the FEBv5 (in black). The deadtime fraction estimated by the ratio between the busy trigger rate and the total trigger rate (filled area) is compared with that obtained from an exponential fit (dots and solid line). All the three sources are shown: the random generator (dots) and the NSB source (triangles) measurements are shown. The vertical gray line at 7~kHz shows the deadtime reduction from 5\% to 0.55\%.}
    \label{fig:deadtime}
\end{figure}

Exponential fitting was performed at different illumination intensities, resulting in an overall minimum deadtime between two consecutive events of $\delta = 744.89 \pm 1.02$ ns, determined through a weighted average across all sources. 
However, when evaluating the camera's deadtime performance, the more meaningful figure of merit is the deadtime fraction. The deadtime fraction is calculated by dividing the aforementioned minimum deadtime by the trigger rate obtained from the exponential fit (solid lines in \autoref{fig:deadtime}). A comparison was made with the deadtime fraction obtained by calculating the ratio between the busy trigger rate (i.e., events triggered while the camera is busy writing data) and the total trigger rate (represented by filled colored bands). Slight discrepancies between the two methods were observed at high trigger rates. This can be attributed to the fact that the exponential distribution is no longer a good approximation of the distribution of the time difference between consecutive events at high rates, resulting in an underestimation of the deadtime percentage. The increase in the deadtime percentage with the second method corresponds to the actual deadtime and is a result of limitations imposed by FIFOs in the DAQ system.
Nevertheless, the two results exhibited agreement up to a few tens of kHz, indicating a nine-fold improvement  compared to the previous FEB version (FEBv5, illustrated in black in \autoref{fig:deadtime}) at 7~kHz.


\begin{figure}[t]
    \centering
    \includegraphics[width=0.9\columnwidth]{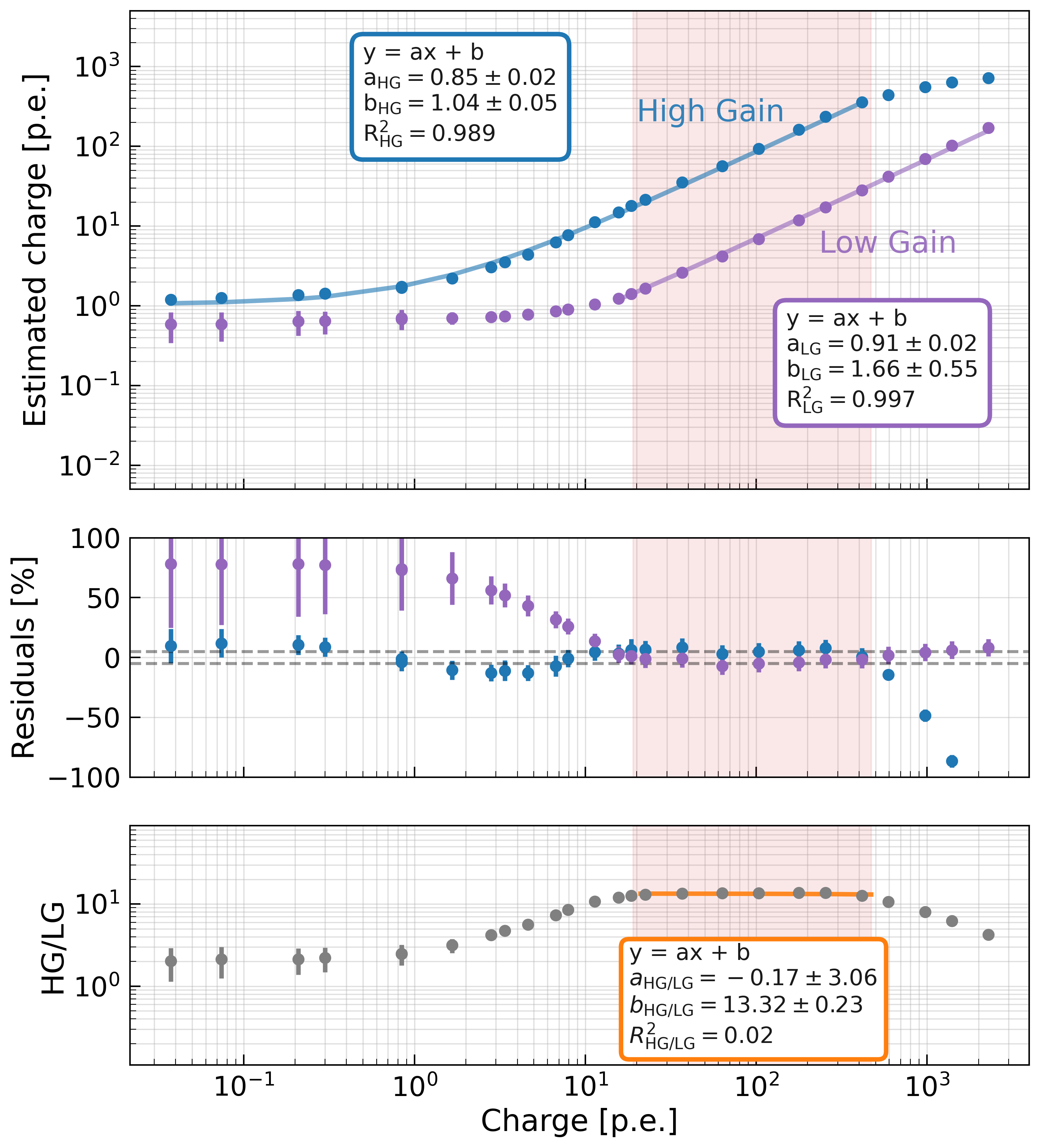}
    \caption{Linearity of 10 FEBv6 modules. The top frame shows the measured charge as a function of the input pulse intensity for the high (blue points) and low (violet points) gain. The two linear fits with the corresponding parameters are shown. The fit residuals are displayed in the middle panel. The bottom panel shows the ratio between the two gains, and the red area indicates the overlapping range between the two gain channels of $\sim20$--$500$~p.e.}
    \label{fig:linearity}
\end{figure}

\section{Linearity}
In this section, we present the results of the linearity test conducted on the 10 FEBv6 devices, demonstrating the distortion of the output as the incident light intensity increases at a specific gain setting. To measure the linearity of the FEBv6 readouts, the charge deposited in each pixel was recorded while illuminating them with the filtered output of the FFCLS. Various \textit{Edmund} filters~\footnote{\url{https://cdn.coverstand.com/30093/556052/1e10501890efd93e0b9a5b8a644dd99d07683283.pdf}} were utilized to achieve an illumination range of 0.1 to 3000 photoelectrons (p.e.).

The deposited charge was obtained by integrating the PMT waveform within a 16 ns readout window centered around the main peak, after subtracting the baseline. The results are presented in \autoref{fig:linearity}. In the top panel, the mean deposited charge is plotted against the true charge for both the high gain channel (blue) and the low gain channel (violet) after conversion to p.e. units. For both channels, a linear least square fit has been performed and the resulting slopes are compatible within 2 standard deviations. 

As shown by the middle panel, the linearity is better than 5\% (dashed gray lines) within 1 standard deviation for the high gain channel between 0.04 and 400~p.e., and for the low gain channel between 10 and 2000~p.e. Thus, the overall dynamic range of this readout spans over 3 decades. The high to low gain ratio remains constant at $\sim$13.32 between $\sim$20 and $\sim$500~p.e. (see bottom panel of \autoref{fig:linearity}).

\section{Conclusions}
We have presented the improved performance of 10 new FEBv6 modules designed for the NectarCAM camera in CTA-North. The integration of an upgraded NECTAr chip operating in "ping-pong" mode has significantly reduced NectarCAM's deadtime fraction by an order of magnitude compared to the previous version, from $5.2\%$ to $0.5\%$ at 7~kHz. This reduction enables higher trigger rates and lowers the energy threshold to approximately 50 GeV at the trigger rate of 7~kHz.

Verification tests were conducted on the 10 new FEBv6 modules with the upgraded NECTAr chips. Pixel timing precision measurements demonstrated precision below 1 ns for incident light intensities above approximately 20 photons. The linearity test of the FEBv6 readouts showed good linearity within 5\% deviation over a wide range of incident light intensities, spanning from 0.04 to 2000 p.e., resulting in a readout dynamic range greater than 3 decades. The deadtime fraction was measured to be 0.55\% at 7 kHz, representing a nine-fold improvement compared to the previous FEB version.

\section*{Acknowledgments}
This work was conducted in the context of the CTA Consortium. We gratefully acknowledge financial support from the agencies and organizations listed here: \url{https://www.cta-observatory.org/consortium_acknowledgments/}.
\bibliographystyle{ICRC}
\bibliography{references}

%

\clearpage



\end{document}